\documentclass[12pt]{iopart}

\usepackage{graphicx}
\usepackage{latexsym}
\usepackage{amsfonts}
\usepackage{amssymb}
\usepackage{setstack}

\newcommand{\h}{{\rm H}}
\newcommand{\mg}{{\rm Mg}}
\newcommand{\ti}{{\rm Ti}}
\newcommand{\hyd}{{\rm H}_{2}}
\newcommand{\mgh}{{\rm Mg}{\rm H}_{2}}
\newcommand{\tih}{{\rm Ti}{\rm H}_{2}}
\newcommand{\mgtih}{{\rm Mg}_{x}{\rm Ti}_{(1-x)}{\rm H}_{2}}
\newcommand{\mgtihy}{{\rm Mg}_{x}{\rm Ti}_{(1-x)}{\rm H}_{y}}
\newcommand{\mgti}{{\rm Mg}_{x}{\rm Ti}_{(1-x)}}
\newcommand{\mgtmh}{{\rm Mg}_{x}{\rm TM}_{(1-x)}{\rm H}_{2}}
\newcommand{\mgtm}{{\rm Mg}_{x}{\rm TM}_{(1-x)}}
\newcommand{\compA}{{\rm Mg}_{0.125}{\rm Ti}_{0.875}{\rm H}_{2}}
\newcommand{\compB}{{\rm Mg}_{0.25}{\rm Ti}_{0.75}{\rm H}_{2}}
\newcommand{\compC}{{\rm Mg}_{0.5}{\rm Ti}_{0.5}{\rm H}_{2}}
\newcommand{\compD}{{\rm Mg}_{0.75}{\rm Ti}_{0.25}{\rm H}_{2}}
\newcommand{\compE}{{\rm Mg}_{0.875}{\rm Ti}_{0.125}{\rm H}_{2}}

\begin{document}

\title{First principles modelling of magnesium titanium hydrides}

\author{S\"{u}leyman Er$^1$, Michiel J van Setten$^2$, Gilles A de Wijs$^3$ and Geert Brocks$^1$}
\address{$^1$Computational Materials Science, Faculty of Science and Technology and MESA+
Institute for Nanotechnology, University of Twente, P.O. Box 217, 7500 AE Enschede, The Netherlands.}
\address{$^2$Institut f\"ur Nanotechnologie, Forschungszentrum Karlsruhe, P.O.~Box~3640, D-76021~Karlsruhe, Germany.}
\address{$^3$Electronic Structure of Materials, Institute for Molecules and Materials, Faculty
of Science, Radboud University Nijmegen, Heyendaalseweg 135, 6525 AJ Nijmegen, The Netherlands.}
\ead{g.brocks@tnw.utwente.nl}

\begin{abstract}
Mixing Mg with Ti leads to a hydride $\mgtih$ with markedly improved (de)hydrogenation properties for $x\lesssim 0.8$, as compared to $\mgh$. Optically, thin films of $\mgtih$ have a black appearance, which is remarkable for a hydride material. In this paper we study the structure and stability of $\mgtih$, $x= 0$-$1$ by first-principles calculations at the level of density functional theory. We give evidence for a fluorite to rutile phase transition at a critical composition $x_{\rm c}= 0.8$-$0.9$, which correlates with the experimentally observed sharp decrease in (de)hydrogenation rates at this composition. The densities of states of $\mgtih$ have a peak at the Fermi level, composed of Ti d states. Disorder in the positions of the Ti atoms easily destroys the metallic plasma, however, which suppresses the optical reflection. Interband transitions result in a featureless optical absorption over a large energy range, causing the black appearance of $\mgtih$.
\end{abstract}
\pacs{71.20.Be, 71.15.Nc, 61.66.Dk, 61.50.Lt}

\submitto{\JPCM}

\maketitle

\section{Introduction}
Hydrogen promises to be a good candidate to replace carbon based fuels in the future \cite{coontz2004nss}. The current inability to store hydrogen in a safe and sufficiently dense form obstructs its use in vehicles or in portable applications. A practical means of storage should yield a system that operates at moderate temperatures and releases hydrogen fast enough to feed a fuel cell \cite{zuttel2003mhs}. Storage should be reversible and the dehydrogenated system should adsorb hydrogen sufficiently fast. Typical target numbers for gravimetric and volumetric hydrogen densities in storage systems are $6$ weight $\%$ (wt\%) and $0.045$ kg/L.

The highest volumetric hydrogen densities among all possible storage forms are achieved in metal hydrides, where atomic hydrogen is bonded in the crystal lattice of bulk metals or alloys \cite{schlapbach2001hsm, zuttel2004hsm}. Reasonable gravimetric hydrogen densities can be obtained if lightweight metals are used. For instance, the simple dihydride $\mgh$ has a gravimetric hydrogen density of $7.7$ wt\%. However, $\mgh$ is so stable that releasing hydrogen at a pressure of 1 bar requires a temperature of around 300$^{\rm o}$C \cite{huot2001mam}, which is impractically high for use in combination with PEM fuel cells. Moreover, the (de)hydrogenation kinetics of $\mgh$ is poor. Surface processes can hamper the kinetics, as the dissociation rate of hydrogen molecules on a Mg surface is small \cite{schlapbach1992hic}, and surface oxidation leads to the formation of a hydrogen diffusion barrier \cite{manchester1988mai}. The surface kinetics can be improved by annealing \cite{zaluska1999nmh} and by adding a catalyst \cite{zaluska2001sca}. Experimental studies also claim that the (de)hydrogenation kinetics of bulk $\mgh$ is very slow. It has been suggested that the kinetics is hampered by the rutile crystal structure of $\alpha$-$\mgh$ \cite{niessen2005ehs,niessen2005hst,vermeulen2006hsm}.

It is well known that 3d transition metals (TMs) act as catalysts to improve the surface kinetics of hydrogen adsorption in Mg \cite{liang1999cet}. To obtain this improvement usually only a few wt\% of TM is added to Mg. Recently it has been shown that also the bulk (de)hydrogenation kinetics can be improved markedly by adding TMs. One has to add a substantially larger amount, however, and make alloys $\mgtm$ with TM $=$ Sc, Ti and $ x\lesssim 0.8$ \cite{niessen2005ehs,niessen2005hst,vermeulen2006hsm}. The hydrides of these alloys have a cubic crystal structure quite unlike the $\alpha$-$\mgh$ rutile structure \cite{latroche2006csm,borsa2007soa,magusin2008hsa,vermeulen2008jmc}. Obviously, to preserve a high gravimetric hydrogen density it is essential to use lightweight TMs. As Sc is too expensive to be used on a large scale, Ti is then the obvious choice. Thin $\mgti$ films, prepared by various experimental techniques, can be reversibly hydrogenated and dehydrogenated \cite{vermeulen2006hsm,vermeulen2006edt,borsa2006mth,slaman2007foh}. The (de)hydrogenation kinetics is fast and increases gradually for alloys with increasing Mg content $x$, until a maximum at $x\approx 0.8$. If the Mg content is larger than this value, the kinetics becomes much slower.

Recent experimental studies reveal that Mg-Ti-H thin films have interesting optical properties, which may lead to applications of these compounds other than hydrogen storage \cite{borsa2006mth,slaman2007foh,baldi2008mth}. In the dehydrogenated state the films are highly reflective, whereas upon hydrogenation they become black, i.e. they have a low reflection and high absorption for light in the visible range. This would allow them to be used as switchable smart coatings for solar panels, or as hydrogen sensors to detect the presence of hydrogen gas, for instance.

Despite the experimental interest in Mg-Ti hydrides, obtaining structural data has proven to be difficult as so far the bulk compounds have resisted synthesis under normal conditions. The only bulk Mg-Ti-H compound synthesized so far is Mg$_{7}$TiH$_{16}$, obtained by letting the dihydrides $\mgh$ and $\tih$ react under extreme conditions of high pressure ($8$ GPa) and temperature ($873$ K) \cite{kyoi2004ntm, ronnebro2005hsa}. In contrast, thin films of $\mgtih$ can be readily obtained for any composition $x$ by various deposition techniques. The limited amount of data on the structure, the thermodynamic stability and the electronic properties of Mg-Ti-H compounds has motivated the present first-principles computational study on $\mgtih$.

First we consider the simple dihydrides $\mgh$ and $\tih$ and subsequently characterize the properties of $\mgtih$, focusing on the relative stability of possible cubic and rutile crystal structures. Limited first-principles calculations on ordered structures have indicated that a cubic fluorite-like crystal structure of $\mgtmh$ is more stable than the rutile structure for a range of early TMs and $x \lesssim 0.8$ \cite{er2009ths,pauw2008cms}. Here we focus on TM = Ti and study such structures in more detail, including the effects of disorder in the positions of the metal atoms, which is relevant for the experimental thin film studies. Our results suggest that the observed change in kinetics as a function of Mg content, is associated with a change in crystal structure. Finally, we calculate the optical properties of $\mgtih$ and analyze them in terms of the compounds' structure and electronic structure. We show that the optical black state, which makes these compounds unique among the reversible hydrides so far, is an intrinsic property \cite{vansetten2009prb}, unlike similar optical states of other Mg-TM-H thin films, which are caused by multiphase inhomogeneities in thin films \cite{richardson01,lohstroh04}.

\section{Computational Details}
\label{sec:compdet}
Most of the present calculations are performed at the level of the  generalized gradient approximation (GGA) to density functional theory, using the projector augmented wave (PAW) technique and a plane wave basis set \cite{blochl1994paw, kresse1999upp}, as implemented in the Vienna {\it Ab initio} Simulation Package (VASP) \cite{kresse1993aim, kresse1996eis, kresse1996eai}. Standard frozen core PAW potentials are used and the H 1s, Mg 2s, Ti 4s and 3d shells are treated as valence shells. As Ti has a partially filled 3d shell we have considered the possibility of spin polarization and ferromagnetic or antiferromagnetic orderings. Unless otherwise mentioned explicitly, most of the compounds studied turn out to be paramagnetic, however. The main results are obtained using the Perdew Wang 91 (PW91) functional \cite{perdew1992aas}, but we have also performed some tests with the hybrid DFT/Hartree-Fock schemes B3LYP and HSE06 \cite{becke1993dft, heyd2003jcp, paier2006jcp, paier2006jcpErr, paier2007jcp}.

We require that the total energy of each compound is converged to within 1 meV/atom with respect to the plane wave kinetic energy cutoff, which is assured by using a cutoff of $650$ eV. Convergence with respect to the ${\it k}$-point sampling for the Brillouin zone (BZ) integration is tested independently on simple compounds using regular meshes of increasing density. We aim at converging total energies on a scale of 1 meV/atom. This is obtained with the following ${\it k}$-point meshes for the simple unit cells: 12$\times$12$\times$8 for hcp unit cells, 16$\times$16$\times$16 for cubic, fcc or fluorite, unit cells and 12$\times$12$\times$16 for bct or rutile unit cells. In studying alloys and their hydrides we construct supercells of these simple unit cells and keep the same ${\it k}$-point grid. Thus employing the equivalent k-points method one avoids relative ${\it k}$-point sampling errors and maintains the accuracy for all cell sizes \cite{froyen1989bzi}.

The molar volume of $\ti$ ($10.64$ cm$^{3}$) is much smaller than that of $\mg$ ($14.00$ cm$^{3}$), which makes finding the optimal cell parameters for $\mgtih$ non-trivial. We start by keeping the atomic fractional coordinates and the {\it c/a} ratio fixed and first carry out volume scans for each composition $x$ and crystal type individually. Values for the lattice parameters are then estimated from the total energy minima. These are then used as starting values for a full optimization of each composition and crystal type.

The cell parameters, including the cell volume, and the atomic positions within the cells are optimized by minimizing the forces and stresses with the conjugate gradient algorithm \cite{payne1992imt}. During optimization Methfessel-Paxton smearing is used with a smearing parameter of $0.1$ eV \cite{methfessel1989hps}. The criteria for self-consistency are set to 10$^{-5}$ eV and 10$^{-4}$ eV for the total energy differences between two consecutive electronic and ionic steps, respectively. Structural relaxation is assumed to be complete if all the forces acting on atoms are smaller than $1$ meV/\upshape{\AA} in the simple cell studies, and $10$ meV/\upshape{\AA} in the supercell studies. After the structures are optimized, the total energies are recalculated self-consistently with the tetrahedron method \cite{blochl1994itm}. The latter technique is also used to calculate the electronic density of states (DOS).

The properties of the $\hyd$ molecule are calculated in a cubic cell with cell parameter 10 \AA, using $\Gamma$-point sampling. The calculated H$-$H bond length, binding energy, and vibrational frequency are $0.748$ \upshape{\AA}, $-4.56$ eV and $4351$ cm$^{-1}$, respectively, in good agreement with the experimental values of $0.741$ \upshape{\AA}, $-4.48$ eV and $4401$ cm$^{-1}$ \cite{huber1979msm}.

To study the stability of $\mgtih$ we look at the reaction
\begin{equation}
	 x\mg  +  (1-x)\ti + \hyd({\rm gas}) \rightarrow \mgtih
	\label{eq:reaction1}
\end{equation}
In principle one should consider the change in Gibbs free energy $G(T,P) = U + PV - TS$ of this reaction. As the thermodynamic properties of hydrogen gas are well documented \cite{hemmes1986jpc}, one can focus upon the solid phases involved in the reaction. In addition, the temperature dependence of $G$ of such solids tends to be relatively small over the temperature range of interest \cite{vansetten2008chm, vansetten2008prb}. A calculation of the change in enthalpy $H=U+PV$ at $T=0$K should therefore be adequate to assess the relative stability of metal hydrides. Since moreover the $PV$ contribution of solids can be neglected, it suffices to focus upon the change in energy $U$.

$U$ corresponds to the total energy of a system as obtained in a DFT calculation. The positions of the atomic nuclei are then fixed, which corresponds to $T=0$ K for classical particles. However, hydrogen is such a light atom that quantum vibrational energies give contributions to the total energies that are not negligible, even at $T=0$ K. These zero-point energies (ZPEs) associated with atomic vibrations are obtained by solving the eigenvalue problem of the dynamical matrix \cite{kresse1995aif}. Slightly displacing an atom from its equilibrium position creates forces on the other atoms. The dynamical matrix can then be constructed by displacing each atom at a time and finite differencing the forces. We use symmetric displacements $dr=\pm0.01$ \upshape{\AA} for each atom and its three degrees of freedom. These calculations are carried out in supercells to properly account for the spatial range of the dynamical matrix.

The ZPE corrected reaction energies are then calculated as
\begin{equation}
	 \Delta{U^{\rm ZPE}} = \sum_{p}(U_{p} + U^{\rm vib}_{p}) - \sum_{r}(U_{r} + U^{\rm vib}_{r}),
	\label{eq:Gibbs3}
\end{equation}
where $r,p$ indicate the reactants and products of the reaction, respectively, $U$ is the total energy obtained from a DFT calculation and $U^{\rm vib}$ is the vibrational ZPE. As it turns out, the ZPE corrections result in a rather constant shift of the reaction energy for the range of compounds studied here. Therefore, we frequently give the reaction energy as $\Delta U$, calculated without $U^{\rm vib}_{p,r}$ contributions, and mention the ZPE corrections separately.

As we will discuss in Sec.~\ref{sec:structures}, we employ two different strategies to model the structures. The first strategy consists of constructing a set of relatively simple ordered structures, inspired by the experimental structure of Mg$_7$TiH$_{16}$ \cite{kyoi2004ntm, ronnebro2005hsa}. The structure of Mg$_x$Ti$_{(1-x)}$H$_y$ as deposited in thin films is cubic with Mg and Ti atoms at fcc positions, but without a regular ordering of Mg and Ti atoms at these positions \cite{borsa2007soa,vermeulen2008jmc}. To model such disordered structures, our second strategy consists of employing special quasi-random structures (SQSs), which enable to model random alloys in a finite supercell \cite{zunger1990sqs}. At each composition $\mgti$ of interest, the Mg and Ti atoms are distributed such, that their lower order correlation functions are equal to those of a perfect random alloy \cite{ruban2003llr, wolverton1999sro}. Hydrogen atoms are then inserted into similar positions as in the simple structures. We use a set of SQSs for each crystal structure (fluorite or rutile) and composition, with a total number of atoms ranging from 48 to 192. Again all cell parameters are optimized, as well as the positions of all atoms within the cell.

To model the optical response of $\mgtih$ we calculate the dielectric function, consisting of interband and
intraband contributions
\begin{equation}
\varepsilon(\omega)=\varepsilon_{\rm inter}(\omega)+\varepsilon_{\rm intra}(\omega),
\label{dieladdi}
\end{equation}
which are calculated separately. The interband contribution $\varepsilon_{\rm inter}$ is calculated in the independent particle approximation, using DFT eigenvalues and wave functions, i.e. neglecting exciton and local field effects. ${\rm Im}[\varepsilon_{\rm inter}(\omega)]$ is calculated directly via the standard longitudinal expression, and ${\rm Re}[\varepsilon_{\rm inter}(\omega)]$ is obtained by a Kramers-Kronig transform \cite{kresseps,vansetten2007prb,vansetten2009prb}. As optical data on hydrides are usually obtained from micro- and nano-crystalline samples whose crystallites have a significant spread in orientation, we use the directionally averaged dielectric function.

Many metals and metal alloys undergo a metal-insulator transition upon full hydrogenation \cite{huiberts96,richardson01,lohstroh04,vangelderen2000prl,vangelderen2002prb,vansetten2005prb,vansetten2007prb,vansetten2007dop}.
$\mgtih$ however retains a finite DOS at the Fermi level. Hence intraband transitions contribute to the dielectric function. We model $\varepsilon_{\rm intra}(\omega)$ by a standard free electron plasma model, with the plasma frequency calculated from the ${\bf k}$-derivatives of the DFT eigenvalues \cite{harl2007prb,vansetten2009prb}. Once the full dielectric function $\varepsilon(\omega)$ is calculated, optical functions such as the extinction coefficient $\kappa(\omega)$ or the refractive index $n(\omega)$ are obtained via the standard expression
\begin{equation}
\varepsilon(\omega) = [n(\omega)+i\kappa(\omega)]^2
\end{equation}
\begin{figure}
	\centering
		\includegraphics[width=11cm]{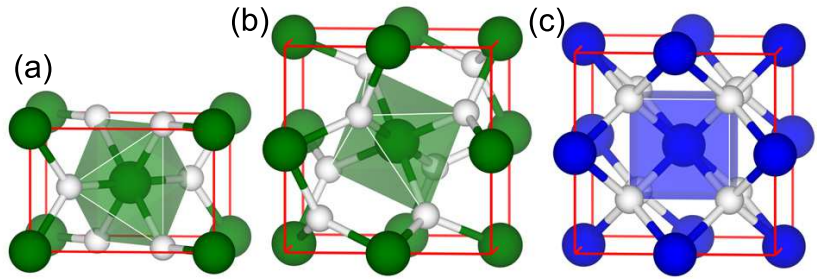}
		\caption{Crystal structures of $\alpha$-$\mgh$ (a), $\beta$-$\mgh$ (b), and $f$-$\tih$ (c). Mg, Ti and H atoms are shown as green, blue and white coloured spheres, respectively; the polyhedra indicate the coordination of the metal atoms by hydrogens. The images are made using the VESTA software \cite{momma2006ccc}.}
	\label{fig:binarydihydrides}
\end{figure}

\section{Simple dihydrides}
\label{sec:simpleH}
As discussed in the introduction, experimental information on the structure of Mg$_{x}$Ti$_{(1-x)}$H$_{2}$ is limited, in particular concerning the positions of the hydrogen atoms. A sensible way to approach the problem is to assume that the structure is similar either to $\mgh$ or to $\tih$. In this section we consider these simple dihydrides in more detail.

\begin{table}
\caption{Calculated and experimental structural parameters of simple metals and hydrides.}
\label{table:BH-structure}
\begin{indented}
\item[]\begin{tabular}{lllllll}
\br
                                     & Space group                          &       &                &       &       &        \\
Compound                             & Unit cell (\upshape{\AA})            &       &                &  x    & y     &  z     \\
\mr
$\mg$                                & P6$_{3}$/mmc (194)                          & $\mg$ &  2c & 1/3   & 2/3   & 1/4       \\
                     & $a = 3.190$ ; $c = 5.180$     &       &              &       &       &           \\
 exp.\cite{villars1991psh} & $a = 3.209$ ; $c = 5.210$     & $\mg$ &  2c & 1/3   & 2/3   & 1/4       \\
$\ti$                                & P6$_{3}$/mmc (194)                          & $\ti$ &  2c & 1/3   & 2/3   & 1/4       \\
                      & $a = 2.916$ ; $c = 4.631$     &       &              &       &       &           \\
 exp.\cite{villars1991psh} & $a = 2.951$ ; $c = 4.686$     & $\ti$ &  2c & 1/3   & 2/3   & 1/4       \\
$\alpha$-$\mgh$                      & P4$_{2}$/mnm (136)                          & $\mg$ &  2a & 0     & 0     & 0         \\
                      & $a = 4.494$ ; $c = 3.005$     & $\h$  &  4f & 0.3044 & 0.3044 & 0         \\
   exp.\cite{bortz1999shp}              & $a = 4.501$ ; $c = 3.010$     & $\mg$ &  2a & 0     & 0     & 0         \\
                      &                                             & $\h$  &  4f & 0.3044 & 0.3044 & 0         \\
$\beta$-$\mgh$                       & Pa$\overline{3}$ (205)                      & $\mg$ &  4a & 0     & 0     & 0         \\
                      & $a = 4.796$                          & $\h$  &  8c & 0.3464 & 0.3464 & 0.3464     \\
$f$-$\tih$                      & Fm$\overline{3}$m (225)                     & $\ti$ &  4a & 0     & 0     & 0         \\
                      & $a = 4.424$                          & $\h$  &  8c & 1/4   & 1/4   & 1/4       \\
   exp.\cite{villars1991psh}              & $a = 4.454$                          & $\ti$ &  4a & 0     & 0     & 0         \\
                     &                                             & $\h$  &  8c & 1/4   & 1/4   & 1/4       \\
$r$-$\tih$                       & P4$_{2}$/mnm (136)                          & $\ti$ &  2a & 0     & 0     & 0         \\
                                     & $a = 3.991$ ; $c = 3.080$     & $\h$  &  4f & 0.322 & 0.322 & 0         \\
\br
	\end{tabular}
\end{indented}
\end{table}

At standard temperature and pressure $\mgh$ has the rutile structure, space group P4$_{2}$/mnm ($136$), see figure \ref{fig:binarydihydrides}. In this so-called $\alpha$-$\mgh$ phase, six hydrogen atoms surround each Mg atom in a distorted octahedron. The distance between Mg and two out of the six surrounding H atoms is $1.938$ \upshape{\AA}, and between Mg and the other four H atoms it is $1.951$ \upshape{\AA}. Each H atom is in the centre of a triangle with Mg atoms at the vertices. Experimentally it has been shown that an increase in the temperature and pressure to $923$ K and $4$ GPa, converts $\alpha$-$\mgh$ to a cubic form, called $\beta$-$\mgh$, with no further information about the positions of the H atoms \cite{bastide1980pmh}. DFT calculations have suggested a structure for $\beta$-$\mgh$ \cite{vajeeston2002pis}. Starting from a fluorite cubic structure with the hydrogen atoms on tetrahedral interstitial sites, space group Fm$\overline{3}$m ($225$), structural optimization gives a cubic structure with reduced symmetry, space group Pa$\overline{3}$ ($205$). This is accompanied by a decrease in the total energy of $0.24$ eV/f.u. and a volume expansion of $\sim 4$\%. Each hydrogen atom is displaced from the centre of a tetrahedron to a base plane, where the coordination by Mg atoms is triangular with the three Mg atoms at a distance $d_1=1.911$ \AA, similar to the $\alpha$-phase. The fourth Mg atom of the original tetrahedron is at a distance $d_2=2.878$ \AA. Each Mg atom is octahedrally coordinated by six H atoms at a distance $d_1$. An additional two H atoms are at $d_2$. At zero pressure the optimized $\beta$-$\mgh$ structure is energetically unfavourable compared to the $\alpha$-$\mgh$ structure by  $0.10$ eV/f.u. and its volume is $\sim 9$\% smaller. The optimized structural parameters obtained in our calculations are given in table \ref{table:BH-structure}. They are in good agreement with available experimental data.

The most stable phase of $\tih$ at standard temperature and pressure has the fluorite structure, space group Fm$\overline{3}$m ($225$), see figure \ref{fig:binarydihydrides}, where the Ti atoms occupy the fcc lattice positions and hydrogen atoms reside at the tetrahedral interstitial positions. The calculated value of the $\ti$$-$$\h$ bond length is $1.916$ \upshape{\AA}. We call this the $f$-$\tih$ structure and to have a comparison between the two structural types of dihydrides we also construct a rutile form of $\tih$, and call it $r$-$\tih$ in table \ref{table:BH-structure}. This artificial rutile structure does not exist in nature to our knowledge. It has a 13\% larger volume than $f$-$\tih$, and its total energy is $0.65$ eV/f.u. higher.


\begin{table}
\caption{Calculated formation energies in eV without ($\Delta{U}$) and with ($\Delta{U^{\rm ZPE}}$) vibrational zero point energy corrections, cf. (\ref{eq:Gibbs3}), compared to experimental values ($\Delta{U^{\rm exp}}$). The latter are taken from \cite{manchester2000pdb} unless indicated otherwise.}
\label{table:BH-energy}
\begin{indented}
\item[]\begin{tabular}{llll}
\br
Compound & $\Delta{U}$ & $\Delta{U^{\rm ZPE}}$ & $\Delta{U^{\rm exp}}$ \\
\mr
$\alpha$-$\mgh$ &  $-0.657$    &   $-0.558$   & $-0.682$ $-0.694$ $-0.771$  \\
 & & & $-0.772$ \cite{bogdanovic1999tim} $-0.779$ \cite{lide2002chc} \\
$\alpha$-$\tih$ &  $-1.467$    &   $-1.279$   & $-1.297$ $-1.351$ $-1.378$ \\
 & & & $-1.450$ \cite{griessen1988tap} $-1.496$  \\
\br
\end{tabular}
\end{indented}
\end{table}

\begin{table}
\caption{Calculated formation energies of $\alpha$-$\mgh$ in eV, obtained with different PAW pseudopotentials for Mg and different DFT functionals or hybrid DFT/Hartree-Fock schemes. The ZPE correction has been calculated at the PW91 level only.}
\label{table:hybrid}
\begin{indented}
\item[]\begin{tabular}{llll}
\br
Mg valence & functional & $\Delta{U}$ & $\Delta{U^{\rm ZPE}}$ \\
\mr
3s         & PW91       & $-0.657$    & $-0.558$       \\
2p3s       & PW91       & $-0.648$    & $-0.549$       \\
3s         & HSE06      & $-0.25$     & $-0.15$        \\
3s         & B3LYP      & $-0.95$     & $-0.85$        \\
\br
\end{tabular}
\end{indented}
\end{table}

The formation energies of the simple hydrides are defined as the reaction energies corresponding to (\ref{eq:reaction1}) for $x=0,1$. The calculated values with and without vibrational ZPE corrections are given in table \ref{table:BH-energy}. ZPEs tend to destabilize the metal hydrides. By far the biggest contribution stems from the difference between the vibrational frequencies of the hydrogen atoms in the metal hydride and in the gas phase $\hyd$ molecules. The ZPEs of the metals are small, i.e. $0.031$ and $0.033$ eV/atom for $\mg$ and $\ti$, respectively (calculated with a 3$\times$3$\times$2 hcp supercell). In general, the more compact the metal hydride structure, the less space for the hydrogen atoms to vibrate, and the higher their vibrational frequencies and the ZPE are. The volumes per formula unit are 30.34 \upshape{\AA}$^{3}$ for $\alpha$-$\mgh$ and 21.64 \upshape{\AA}$^{3}$ for $f$-$\tih$. The calculated ZPEs are  $0.406$ eV/f.u. and $0.501$ eV/f.u., respectively. These are calculated using a 2$\times$2$\times$3 rutile supercell for $\mgh$, containing 72 atoms, and a 2$\times$2$\times$2 fluorite supercell for $\tih$, containing 96 atoms. The calculated ZPE of a $\hyd$ molecule is $0.276$ eV. Consequently, including ZPE corrections destabilizes $\alpha$-$\mgh$ by approximately 0.1 eV, whereas the more compact $f$-$\tih$ is destabilized by almost 0.2 eV, see table~\ref{table:BH-energy}.

The experimental numbers obtained for the formation energy of $\tih$ range from $-1.30$ to $-1.50$ eV. The calculated numbers are in good agreement with these results. Experimental values of the formation energy of $\mgh$ range from $-0.68$ to $-0.78$ eV. The calculated values overestimate these numbers somewhat, in particular if one includes ZPEs. Such differences between calculated and experimental formation energies are also found in other simple alkali and alkaline earth hydrides and seem to be typical for the use of GGA functionals such as PW91 or PBE. To see whether the calculated values can be improved, we test the use of a hard PAW pseudopotential for Mg, which includes the 2p shell as valence electrons, as well as the use of the hybrid DFT/Hartree-Fock schemes B3LYP and HSE06. The results, listed in table~\ref{table:hybrid}, show that the effect of using a hard PAW potential is minor. Hybrid DFT/Hartree-Fock schemes do not necessarily give an improvement over DFT/GGA functionals. The HSE06 and B3LYP schemes lead to formation energies of $\mgh$ that are respectively much larger and somewhat smaller than the experimental one. Since in the following we are foremost interested in energy differences between crystal structures, we stick to using the PW91 GGA functional.


\begin{figure}
	\centering
		\includegraphics[width=0.7\textwidth]{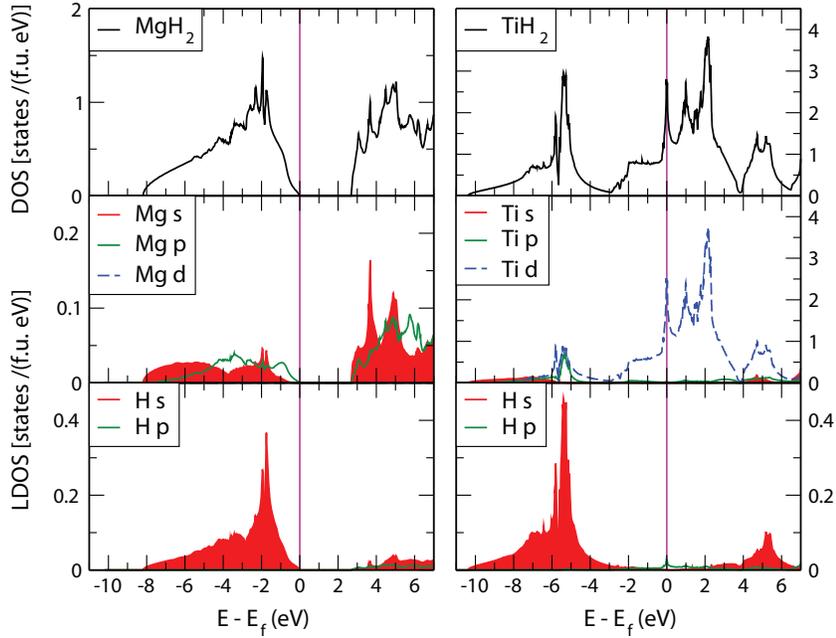}
	  \caption{Electronic density of states (DOS) of $\beta$-$\mgh$ and $f$-$\tih$. The origin of the energy scale is set at the top of the valence band. The top figure gives the total DOS, and the middle and bottom figures represent the local DOS, projected on the atoms; s, p and d contributions are shown in red, green and blue, respectively. The regions under the s curves are shaded and the d curves are dashed.}
	\label{fig:simpleH}
\end{figure}

$\alpha$-$\mgh$ is an insulator with an experimental optical band gap of $5.6$ eV \cite{isidorsson2003opm}. As usual DFT calculations underestimate this gap \cite{vansetten2007prb}; the value obtained for $\alpha$-$\mgh$ is $3.8$ eV, whereas the calculated gap for $\beta$-$\mgh$ is $2.7$ eV. The electronic (projected) densities of states ((P)DOS) of the cubic simple dihydrides $\beta$-$\mgh$ and $f$-$\tih$ are shown in figure~\ref{fig:simpleH}. The fact that many fully hydrided simple metals and alloys are insulators can be understood from simple chemical arguments \cite{vangelderen2000prl,vangelderen2002prb,vansetten2005prb,vansetten2007prb,vansetten2007dop}. The Mg atoms almost fully donate their valence (2s) electrons to the H atoms. Indeed the PDOS in figure~\ref{fig:simpleH} shows that the valence bands have a predominant H 1s character, whereas the Mg 2s states emerge in the conduction bands. In contrast, $\tih$ is metallic with a considerable DOS at the Fermi level. The PDOS shows that the states around the Fermi energy have a dominant Ti 3d character, whereas, as for $\mgh$, the lower valence bands are dominantly H 1s. It suggests that in $\tih$ two electrons per metal atom are transferred to hydrogen and, as Ti atoms have four valence electrons (s$^2$d$^2$), they remain in an open-shell configuration in the hydride. Figure~\ref{fig:simpleH} illustrates that the Ti d electrons participate significantly less than the s electrons to the bonding to H atoms.

\section{$\mgtih$}
\subsection{Structures}
\label{sec:structures}

The only detailed structure of $\mgtih$ available from experiment is for the composition $x=0.875$. In the Mg$_7$TiH$_{16}$ high pressure phase the metal atoms are in fcc positions and are ordered as in the Ca$_7$Ge structure \cite{kyoi2004ntm,ronnebro2005hsa}. The H atoms are at interstitial sites, but displaced from their ideal tetrahedral positions. This structure has 96 atoms in the unit cell. We use it as a starting point to construct relatively simple, fluorite-type structures for other compositions $x$. For $x=0.125, 0.875$ we use the Ca$_7$Ge structure to order the metal atoms, for $x=0.25, 0.75$ the Cu$_3$Au ($L1_2$) structure, and for $x=0.5$ the CuAu ($L1_0$) structure. The H atoms are placed at or close to tetrahedral interstitial positions. In each of the structures and compositions the cell parameters are optimized, as well as the positions of all atoms within the cell. Care is taken to allow for breaking the symmetry in the atomic positions, as the hydrogen atoms are often displaced from their ideal tetrahedral positions. Although strictly speaking such structures are then no longer ideal fluorite structures anymore, we still use the term fluorite in the following.

\begin{table}
\caption{Optimized cell parameters and atomic positions of the fluorite-type simple $\mgtih$ structures.}
\label{table:fMgTiH-structure}
\begin{indented}
\item[]\begin{tabular}{llllll}
\br
Compound, Space group                &         &        &         &       &    \\
Unit cell (\upshape{\AA})            &         &        &    x    & y     &  z     \\
\mr
$\compA$                             &  $\mg$  &4{\it a} & 0      & 0      & 0          \\
Fm$\overline{3}$m (225)              & $\ti$1  &4{\it b} & 1/2    & 1/2    & 1/2        \\
{\it a} = 8.8896                     & $\ti$2  &24{\it d}& 0      & 1/4    & 1/4        \\
                                     & $\h$1   &32{\it f}& 0.1285 & 0.1285 & 0.1285 \\
                                     & $\h$2   &32{\it f}& 0.3755 & 0.3755 & 0.3755   \\
$\compB$                             & $\ti$   & 3{\it c}& 0      & 1/2    & 1/2          \\
Pm$\overline{3}$m (221)              & $\mg$   & 1{\it a}& 0      & 0      & 0       \\
{\it a} = 4.4648                     & $\h$    & 8{\it g}& 0.2566 & 0.2566 & 0.2566 \\
$\compC$                             & $\ti$1  & 1{\it a}& 0      & 0      & 0       \\
P4/mmm (123)                         & $\ti$2  & 1{\it c}& 1/2    & 1/2    & 0       \\
{\it a} = 4.4720                     & $\mg$   & 2{\it e}& 0      & 1/2    & 1/2     \\
{\it c} = 4.6544                     & $\h$    & 8{\it r}& 0.25   & 0.25   & 0.2352  \\
$\compD$                             & $\ti$   & 1{\it a}& 0      & 0      & 0       \\
Pm$\overline{3}$m (221)              & $\mg$   & 3{\it c}& 0      & 1/2    & 1/2       \\
{\it a} = 4.6208                     & $\h$    & 8{\it g}& 0.2432 & 0.2432 & 0.2432 \\
$\compE$                             & $\ti$   & 4{\it a}& 0      & 0      & 0       \\
Fm$\overline{3}$m (225)              & $\mg$1  & 4{\it b}& 1/2    & 1/2    & 1/2     \\
{\it a} = 9.3565                     & $\mg$2  &24{\it d}& 0      & 1/4    & 1/4     \\
                                     & $\h$1   &32{\it f}& 0.1194 & 0.1194 & 0.1194   \\
                                     & $\h$2   &32{\it f}& 0.3725 & 0.3725 & 0.3725   \\
$\compE$ exp.\cite{ronnebro2005hsa}        & $\h$1   &32{\it f}& 0.094(2) & 0.094(2) & 0.094(2)   \\
{\it a} = 9.564(2)                   & $\h$2   &32{\it f}& 0.365(2) & 0.365(2) & 0.365(2)   \\
\br
	\end{tabular}
\end{indented}
\end{table}

\begin{figure}
	\centering
		\includegraphics[width=11cm]{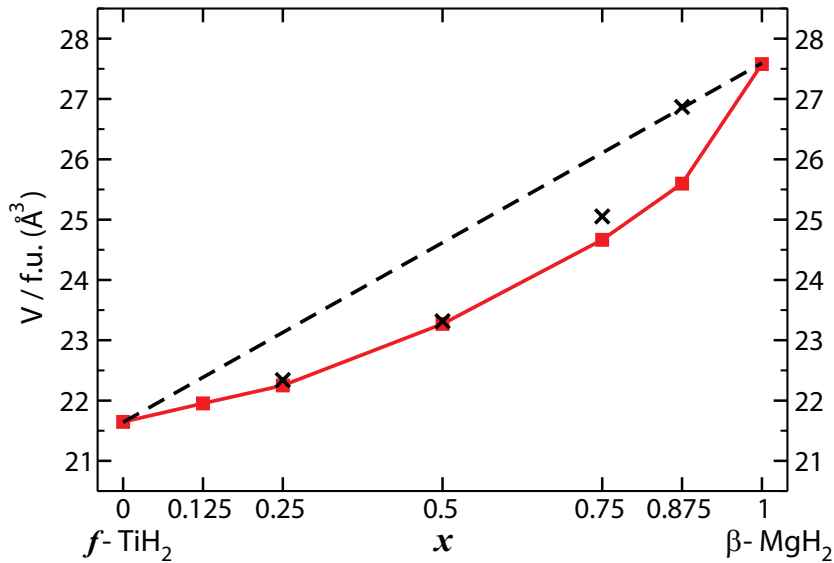}
		\caption{The volumes per formula unit (V/f.u) in \AA$^3$ of $\mgtih$ in the ordered fluorite structures, as a function of the composition $x$. The dashed line marks a linear interpolation between $f$-$\tih$ and $\beta$-$\mgh$. The crosses mark the volumes of the special quasi-random structures.}
	\label{fig:Volfluorite}
\end{figure}

The optimized fluorite structures are given in table~\ref{table:fMgTiH-structure}. Most of these structures are cubic, except $\compC$, which is tetragonally distorted. Comparing the calculated and the experimental structures of $\compE$ one observes that the calculated lattice parameter is $2$\% smaller than the experimental one. Although such a result is not uncommon for DFT calculations, note that the differences between calculated and experimental lattice parameters of $\mgh$ and $\tih$ are much smaller, see table~\ref{table:BH-structure}. The Wyckoff positions and the hydrogen-metal coordinations of the calculated and experimental structures of $\compE$ agree, but the absolute difference between the calculated and experimental H positions is 0.1-0.4 \AA. This is not a volume effect, since optimizing the structure using the experimental or the calculated lattice parameter gives the same H positions. We have also checked that breaking the symmetry by random displacement of the H atoms yields the same optimized Fm$\overline{3}$m (225) structure. From our calculations the experimental structure is $0.766$ eV/f.u. higher in energy.

The calculated tetrahedral environment of a H1 atom in $\compE$ consists of a Ti atom at a distance of 1.935 \AA\ and three Mg2 atoms at 2.058 \AA; the experimental distances are 1.557 \AA\ and 2.293 \AA, respectively. The calculated tetrahedral environment of a H2 atom consists of a Mg1 atom at 2.066 \AA\ and three Mg2 atoms at 2.013 \AA\, compared to experimental distances of 2.236 \AA\ and 2.021 \AA. The calculated H$-$Ti distance is close to the 1.916 \AA\ found in $\tih$, see table~\ref{table:BH-structure}, whereas the experimental distance is significantly smaller. In fact, for all compositions in table~\ref{table:fMgTiH-structure} we find H$-$Ti distances in the range 1.91-1.94 \AA. From the calculated H$-$Mg distances one observes that the H atoms are closer to ideal tetrahedral positions than in the experimental structure. From the positions of the H2 atoms it is clear, however, that both the calculated and the experimental structure are closer to an ideal fluorite structure than to $\beta$-$\mgh$.

One should note that the experimental structure of Mg$_7$TiH$_{16}$ is based upon room temperature XRD data. As XRD measures the electronic charge distribution, modelling the data with atomic scattering functions introduces errors on the positions of light atoms such as hydrogen. In particular, the asymmetry in the hydrogen positions is sometimes exaggerated. In that case low temperature neutron diffraction can give a structure that is much closer to the calculated structure \cite{hartman2007jssc,vansetten2008chm}.

The calculated volumes of the fluorite structures of $\mgtih$ as a function of the composition $x$, normalized per formula unit, are shown in figure~\ref{fig:Volfluorite}. According to Zen's law of additive volumes one would expect a linear dependence \cite{hafner1985nvs}. The curves deviate slightly, but distinctly, from straight lines, with a maximum deviation of 5.4~\% at $x=0.5$. This deviation is consistent with the experimental observations on $\mgtih$ \cite{vermeulen2006hsm,borsa2007soa}. It is also observed in simple metal alloys \cite{hafner1985nvs}.

\begin{figure}
		\includegraphics[width=16cm]{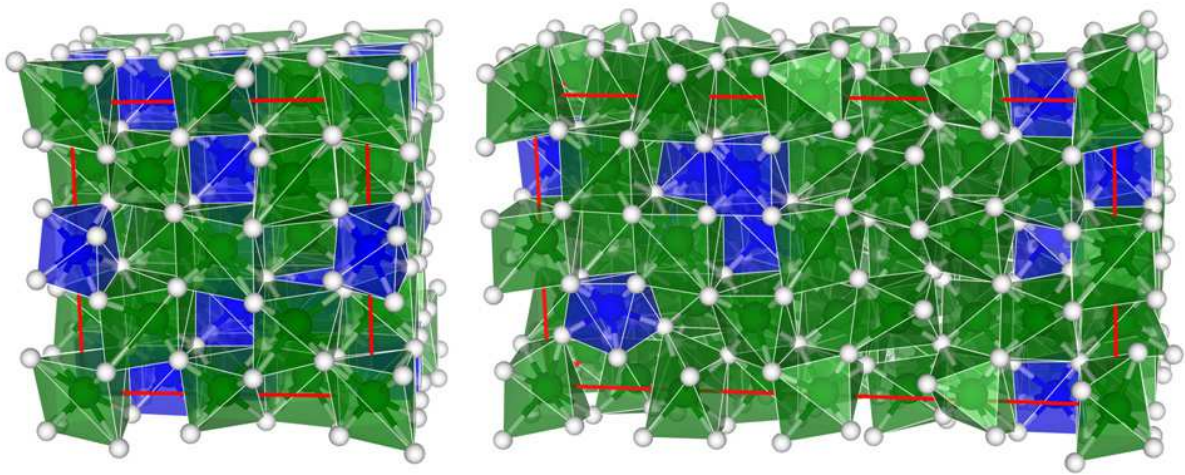}
		\caption{Cells to model special quasi-random structures of $\mgtih$ for $x = 0.75$ (left image) and $x = 0.875$ (right image) compositions. Mg, Ti and H atoms are shown as green, blue and white coloured spheres, respectively; the polyhedra indicate the coordination of the metal atoms by hydrogens \cite{momma2006ccc}.}
		\label{fig:fsqs}
\end{figure}

The binary phase diagram of Mg and Ti indicates that they do not form a stable bulk alloy. It is however claimed that synthesis by mechanical alloying may be possible \cite{liang2003smt}. In addition, thin films of $\mgti$ are readily made, which can be reversibly hydrogenated \cite{niessen2005ehs,vermeulen2006hsm,vermeulen2006edt,borsa2006mth,borsa2007soa,slaman2007foh}. The crystal structure of $\mgtihy$ in thin films, $x\lesssim 0.8$, $y\approx 1$-2, is cubic, with the Mg and Ti atoms at fcc positions, but no detectable regular ordering of Mg and Ti
atoms at these positions \cite{borsa2007soa,vermeulen2008jmc}. The positions of the H atoms have not been determined from experiment.

To model such disordered structures, we perform calculations on SQSs, which enable to model random alloys in a finite supercell. We use a 32 atom supercell to model SQSs of fcc Mg$_x$Ti$_{1-x}$ for $x=0.25,$ $0.5,$ and $0.75$, and a 64 atom supercell for $0.875$ \cite{ruban2003llr}. Inserting hydrogen atoms in tetrahedral interstitial positions then gives supercells with a total number of atoms of 96 and 192, respectively. Optimizing the structure leads to some displacements in the positions of the metal atoms, but they remain close to fcc. There is a larger spread in the positions of the H atoms, which increases with increasing $x$.
Representative examples of SQSs are given in figure~\ref{fig:fsqs}. Whereas in the ordered structures of table~\ref{table:fMgTiH-structure} the nearest neighbour Ti$-$Ti distance increases with increasing Mg content $x$, in the SQSs one can always find at least one Ti$-$Ti pair at a distance comparable to that in $f$-$\tih$, for the compositions studied.

\begin{figure}
	\centering
		\includegraphics[width=11cm]{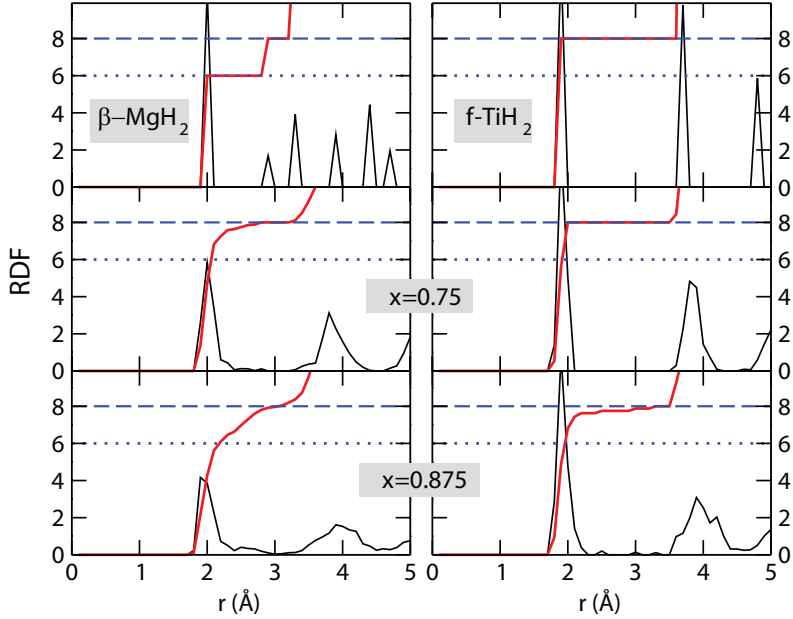}
		\caption{Radial distribution functions (RDFs) for Mg$-$H (left column) and Ti$-$H (right column) pairs in $\mgtih$, $x=0.75$, $0.875$ fluorite-type SQSs, compared to the RDFs of $\beta$-$\mgh$ and $f$-$\tih$. The red lines indicate the integrated RDF. Horizontal dotted and dashed lines indicate the H coordination numbers of Mg and Ti in the simple dihydrides, respectively.}
	\label{fig:rdf}
\end{figure}

The optimized volumes of the SQSs are shown in figure~\ref{fig:Volfluorite}. For $x\leq 0.75$ these volumes are within 1.5\% of those of the ordered structures. Remarkably, the SQS volume of $\compE$ is 4.9\% larger than that of the ordered structure of table~\ref{table:fMgTiH-structure}. This is indicative of a structural change at this composition, which can be clarified by calculating the radial distribution functions (RDFs) of the SQSs. The RDFs of $\mgtih$ for $x = 0.75, 0.875$ are plotted in figure~\ref{fig:rdf}. Shown are the RDFs of Ti$-$H and Mg$-$H distances and the corresponding RDFs in the cubic $\beta$-$\mgh$ and $f$-$\tih$ compounds. Integrating the RDF gives the number of hydrogen atoms surrounding each metal atom. The number in the first shell then corresponds to the coordination number of the metal atom.

In $f$-$\tih$ each Ti atom is coordinated by eight H atoms at a distance of approximately $1.9$ \upshape{\AA}, whereas in the Mg atoms in $\beta$-$\mgh$ are coordinated by six H atoms at $2.0$ \AA, and by an additional two H atoms at $2.9$ \AA. The RDFs of the SQSs show that in $\mgtih$ each Ti is coordinated as in $f$-$\tih$, irrespective of the composition $x$. For $x=0.75$ each Mg atom is also coordinated by eight atoms at a similar distance, as in a fluorite structure. However, for $x=0.875$ the RDF for Mg$-$H starts to resembles that of $\beta$-$\mgh$, i.e. a peak representing six H atoms at $\sim 2.0$ \AA, and a shoulder representing two H atoms at $\sim 2.9$ \AA. Apparently a large Mg content is required before such a structure can develop, which is accompanied by a volume expansion with respect to an ideal fluorite structure. In order not to complicate the terminology we call all such structures fluorite-type in the following.

Since we will compare the stability of these fluorite-type structures to rutile-type structures $\mgtih$, we also make models of the latter using the $\alpha$-MgH$_2$ structure as a starting point. To construct simple, ordered structures, we replace a fraction $x$ of the Mg atoms by TM atoms and use the smallest supercell of the rutile structure where this leads to an integer number of atoms. In case of multiple possible cells, the results given below refer to the cell that leads to the lowest energy. Rutile SQSs of $\mgti$ for $x=0.25$-$0.875$ are constructed using standard algorithms \cite{ruban2003llr}. As it turns out that rutile structures are relatively unstable over a large composition range, we refrain from giving their structural details here.

\subsection{Stability}

Experimentally it is concluded that Mg and Ti do not form a thermodynamically stable bulk alloy. Indeed we find that its costs energy to make alloys from the bulk metals Mg and Ti. For instance, the energy costs of making $\mgti$ fcc SQSs from bulk hcp Mg and Ti are 0.16 and 0.11 eV/atom for $x=0.5$ and $0.75$, respectively, and for $\mgti$ in the simple, ordered CuAu and Cu$_3$Au structures the energy costs are 0.22 and 0.15 eV/atom, respectively. Calculations on hcp $\mgti$ decrease these numbers by $\sim 0.04$ eV, which implies that the hcp alloy is more stable than the fcc alloy, but it is still unstable. Moreover, simple estimates of the configurational entropy show that the alloys cannot be stabilized at a reasonable temperature by entropy. Metastable structures can still be relevant, though, as thin film deposition of $\mgti$ yields a hcp structure that survives a large number of (de)hydrogenation cycles.

\begin{figure}
	\centering
		\includegraphics[width=11cm]{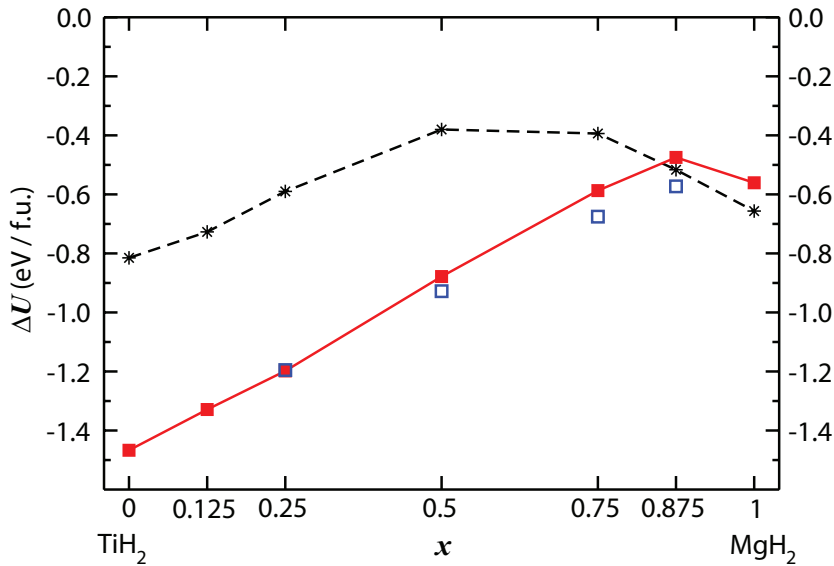}
			  \caption{Formation energies of $\mgtih$ as function of composition $x$ for ordered fluorite-type (filled red squares) and rutile-type (black stars) structures. The lines are added to guide the eye. The open blue squares indicate the formation energies of fluorite-type SQSs.}
	\label{fig:deltaH}
\end{figure}

The formation energy of $\mgtih$ according to (\ref{eq:reaction1}) is given in figure \ref{fig:deltaH} for the fluorite and rutile structures. The numbers are given without ZPE corrections, which we will discuss later. Clearly all numbers are negative, which implies that the compounds are stable with respect to decomposition into the elements. In addition, over most of the composition range the fluorite structure is more stable than the rutile structure. For $x$ larger than a critical composition $x_{\rm c}$ the rutile structure becomes lower in energy. If we use the energies calculated for the simple ordered structures as a starting point and linearly interpolate between the compositions, we find $x_{\rm c}\approx 0.83$. The fact that $x_{\rm c}$ has a high value makes sense, since the energy difference between the $\alpha$-$\mgh$ (rutile) and $\beta$-$\mgh$ (``fluorite'') phases is only 0.10 eV/f.u., whereas between $f$-$\tih$ (fluorite) and $r$-$\tih$ (rutile) it is 0.65 eV/f.u.. Therefore, it requires a high Mg content to force the structure into rutile. Indeed, experimental results suggest that the cross-over from fluorite to rutile structures in $\mgtih$ occurs for $x$ somewhere in the range 0.8-0.9 \cite{borsa2007soa,vermeulen2008jmc}.

Note that for all compositions the structures are metastable with respect to decomposition into $\tih$ and $\mgh$, as is evident from figure \ref{fig:deltaH}. Evidently, the decomposition is kinetically hindered as it does not occur experimentally. Upon dehydrogenation the cubic $\mgtih$ converts into hcp $\mgti$. Although the latter is metastable with respect to Ti and Mg, its decomposition is also kinetically hindered, as thin films can be (de)hydrogenated reversibly. The apparent stability of these compounds is quite remarkable.

The formation energies of the fluorite SQSs are also given in figure \ref{fig:deltaH}. Within the constraints of the SQS approach, some variation in the relative ordering of Ti and Mg atoms is still possible at each composition $x$. We estimate that this variation gives a spread in the energies of $\sim 0.1$ eV/f.u.. From figure~\ref{fig:deltaH} one also observes that the most stable SQSs are slightly more stable than the ordered structures, by up to $\sim 0.1$ eV/f.u. for $x=0.875$. We have also constructed rutile SQSs of $\mgti$ for $x=0.25$-$0.875$. However, upon optimization these structures turn out to be unstable and spontaneously convert into a fluorite-type structure (i.e. using the conjugate gradient algorithm). This would suggest that a fluorite-type structure is stable at least up to a composition $x=0.875$. As the energies of the two types of structures are quite similar over a range of compositions close to the critical composition $x_{\rm c}$, the exact value of $x_{\rm c}$ is difficult to determine. Note that here we only consider simple ordered and SQSs models of rutile and flourite-type lattices. Of course, it cannot be excluded that different structural models describe the true ground state. However, our structures are judicially chosen so that our studies are relevant to understand the experiments; indeed we obtain a good agreement with experimental studies.

The ZPE corrections destabilize the simple hydrides $\alpha$-$\mgh$ and $\alpha$-$\tih$, see table~\ref{table:BH-energy}, and one expects this also to be the case in $\mgtih$. The size of the ZPE correction in the simple hydrides indicates that it decreases with increasing $x$ from $0.19$ eV/f.u. at $x=0$ to $0.10$ eV/f.u. at $x=1$. We suggest that this decrease is approximately linear in $x$. An explicit calculation on the SQS of the $x=0.75$ composition gives a ZPE correction of 0.12 eV/f.u., in agreement with this suggestion.

We now consider to what extend $\mgtih$ is suitable as a hydrogen storage material. Experimentally it has been observed that the dehydrogenation kinetics becomes markedly slower if $x\gtrsim x_0 = 0.8$ . The results of figure~\ref{fig:deltaH} strongly suggest that $x_0 = x_c$, i.e. the composition at which the phase transition between fluorite and rutile structures takes place. Lightweight materials require a high content of magnesium, but to have a stable fluorite structure it should not exceed the critical composition $x_c$. We focus upon the composition $\compD$, which should have the fluorite structure. It has gravimetric and volumetric hydrogen densities of 6.2 wt\% and 0.135 kg/L, respectively. Its calculated formation energy is $-0.59$ eV/f.u. (without ZPE), which means that this compound is slightly less stable than $\alpha$-$\mgh$ with respect to decomposition into the elements. A decreased stability of the hydride should ease the dehydrogenation reaction. However, in order that the process is reversible, it is not advantageous if the alloy $\mgti$ dissociates upon dehydrogenation. If we calculate the hydrogenation energy according to the reaction
\begin{equation}
\label{eq:hydrogenation}
\mgti  +  \hyd(g) \rightarrow  \mgtih,
\end{equation}
we find $-0.76$ eV/f.u. for $x=0.75$. This number is in good agreement with the experimental value of $-0.81$ eV/f.u. of \cite{gremaud2007hoc}, obtained if the thin film correction suggested there is included. This hydrogenation enthalpy is somewhat lower than that of pure Mg, showing that alloying Mg with Ti does not improve the thermodynamics as compared to pure Mg. On the other hand, it is also not much worse, and the kinetics of (de)hydrogenation of $\compD$ is markedly better than that of pure Mg.

\subsection{Electronic structure and optical properties}
\begin{figure}
	\centering
		\includegraphics[width=11cm]{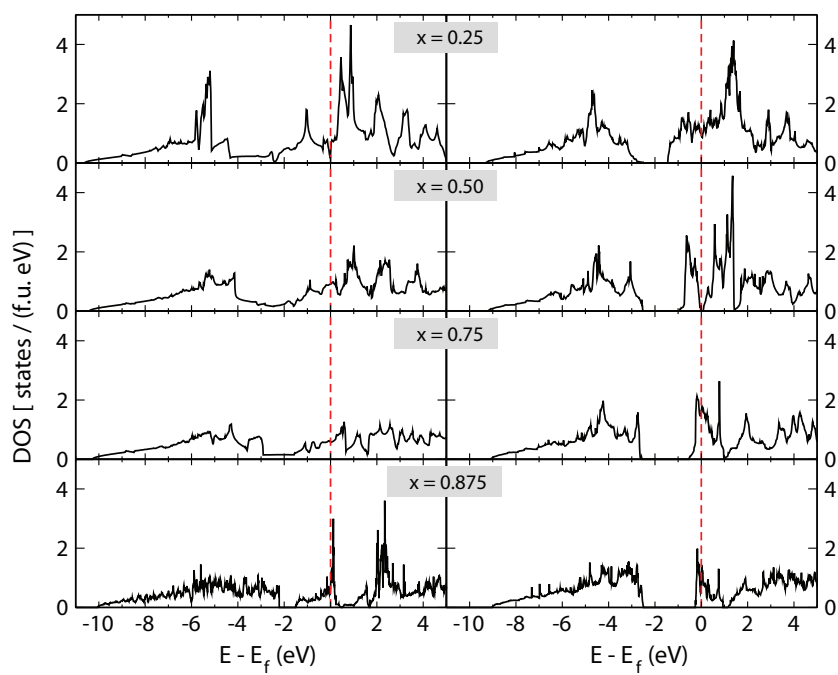}
		\caption{Electronic density of states (DOS) per formula unit for the fluorite (left column) and rutile (right column) $\mgtih$ structures.}
	\label{fig:DOSternary}
\end{figure}

\begin{figure}
	\centering
		\includegraphics[width=11cm]{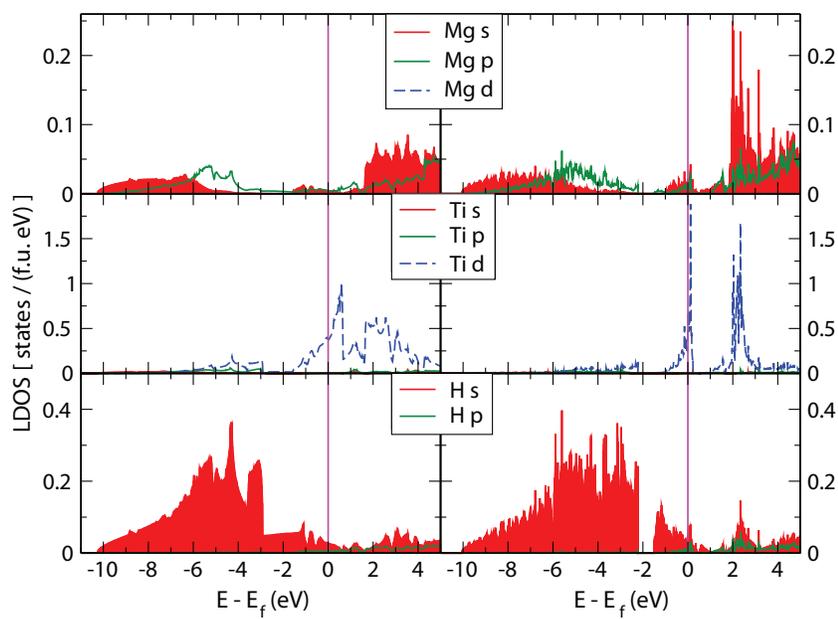}
		\caption{Local density of states (LDOS) for fluorite-$\mgtih$, $x = 0.75$ (left column) and $x = 0.875$ (right column). The origin of the energy scale is set at the Fermi level. Atomic s, p and d contributions are shown in red, green and blue, respectively. The regions under the s curves are shaded and the d curves are dashed.}
	\label{fig:LDOSternary}
\end{figure}

The calculated electronic density of states (DOS) of $\mgtih$ in the simple ordered structures is shown in figure~\ref{fig:DOSternary} for fluorite- and rutile-type structures. Qualitatively these DOSs can be interpreted as linear combinations of DOSs of the simple hydrides, see figure~\ref{fig:simpleH}. In the DOSs of the rutile structure one can still distinguish the band gap that originates from $\mgh$. The states just above the gap originate from Ti, as discussed in Sec.~\ref{sec:simpleH}. In particular the DOS of rutile-$\compE$ can be interpreted as Ti-doped $\alpha$-$\mgh$. Enlarging the Ti content then increasingly fills the gap with Ti-derived states and for rutile-$\compB$, the DOS is dominated by a Ti-derived peak at the Fermi level.

A similar qualitative interpretation of the DOSs also holds for the $\mgtih$ fluorite structures. It is somewhat less clear because the band gap in (fluorite-type) $\beta$-$\mgh$ is smaller than that of (rutile) $\alpha$-$\mgh$. Still the DOS of $\compE$ can clearly be interpreted as Ti-doped $\beta$-$\mgh$, where Ti-derived states appear in the gap. Upon increasing the Ti content, the DOS more and more resembles that of $f$-$\tih$ and the gap is completely filled. The latter has implications for the optical properties, as we will discuss below.

In order to be able to identify the character of the electronic states, the atom projected DOS of the simple structures is shown in figure~\ref{fig:LDOSternary} for fluorite-type $\mgtih$, $x=0.75$, $0.875$, which are the most relevant structures and compositions for studying the optical properties \cite{borsa2007soa}. It is obvious that the peaks at the Fermi level are dominated by states of Ti 3d character, as in $f$-$\tih$, see figure~\ref{fig:simpleH}. For $\compE$ these peaks are narrow, reflecting the large distance of 6.62 \AA\ between neighbouring Ti atoms. One observes the e$_{\rm g}$-t$_{\rm 2g}$ splitting that is typical of d states in a cubic crystal field. In this case the latter originates from the cubic coordination of each Ti atom by H atoms. Local magnetic moments of 1.5 $\mu_{\rm B}$ develop on the Ti atoms and we predict an anti-ferromagnetic ordering of these moments at low temperature. The estimated N{\' e}el temperature is only 40K, however, demonstrating that the moments are atomic like. If we increase the Ti content, the states on neighbouring Ti atoms interact more strongly. For $\compD$ the nearest neighbour distance between Ti atoms is 4.62 \AA. The Ti $d$ derived peaks are much wider, and the local magnetic moments disappear.

The states in the PDOS at low energy are dominantly H 1s derived states, as expected. Interestingly, the states around the Fermi level also have a H contribution. These contributions are spread out over an energy range that is much wider in $\mgtih$ than in the simple hydrides $\mgh$ and $\tih$, see figure~\ref{fig:simpleH}. The spread occurs in H derived bands of s character, as well as in the those of p character. Since transitions between such H derived bands give a large contribution to the optical response of a hydride \cite{vansetten2007prb}, one might  expect the response to be spread over a large energy range.

\begin{figure}
\centering
		\includegraphics[width=11cm]{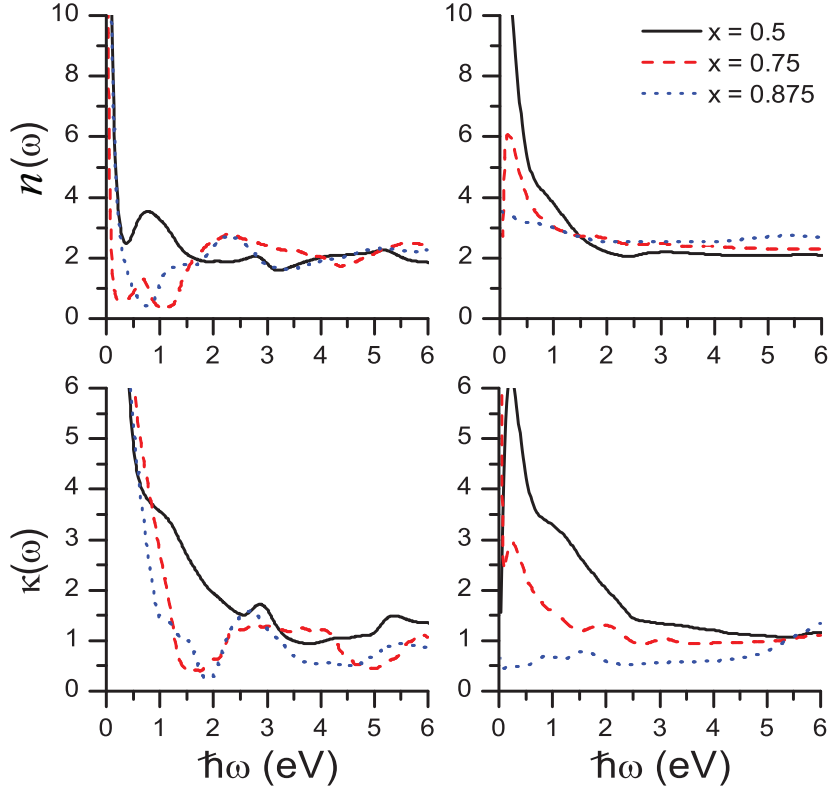}
		\caption{  Index of refraction (top) $n(\omega)$ and extinction coefficient $\kappa(\omega)$ (bottom) of $\mgtih$ as function of frequency $\omega$. The black (solid), red (dashed), and blue (dotted) lines are for the compositions $x=0.5,0.75,0.875$, respectively. The left and right columns are for the ordered structures and the SQSs, respectively.}
		\label{fig:optix}
\end{figure}

We calculate the optical response functions of $\mgtih$ as described in Sec.~\ref{sec:compdet}. The refractive index $n(\omega)$ and the extinction coefficient $\kappa(\omega)$ are shown in figure~\ref{fig:optix} for the compositions $x=0.5$, 0.75, and 0.875. Most relevant for a comparison to the optical experiments on thin films are the data for the SQSs. Comparing our data on the SQSs and the ordered structures, we observe that one main effect of disorder in the positions of the metal atoms, as modelled by a SQS, is smoothing the spectrum.

In addition, disorder strongly affects the lower end of the spectrum, i.e. $\omega=0$-$1.5$ eV, in particular for the Mg-rich compositions $x=0.75,0.875$. The extinction coefficient $\kappa(\omega)$ of the SQSs in this frequency range is smaller than that of the ordered structures and much closer to its value at high frequency. The large $\kappa(\omega<1$~eV) in the ordered structures is typical for the electron plasma response of metals, which matches the DOSs shown in figure~\ref{fig:DOSternary}. As the metallic properties of $\mgtih$ originate from the Ti atoms only, introducing disorder in the Ti positions can destroy the electron plasma, suppressing the large $\kappa(\omega<1$~eV) \cite{vansetten2009prb}. As can be observed in figure~\ref{fig:optix}, compositions with a low Ti content are particularly sensitive to disorder, as the coupling between Ti atoms is relatively small.

Also the structure in the refractive index of the ordered structures at low frequency is suppressed in the SQSs and the refractive index at $\omega > 1$ eV becomes close to constant. Assuming that one may use bulk optical functions to model the optical properties of $\mgtih$, the reflectance (at normal incidence) of a surface is given by the usual Fresnel expression $R=((n-1)^2+\kappa^2)/((n+1)^2+\kappa^2)$. For a metal usually $\kappa \gg n$, giving rise to a large reflectance. In this case both $\kappa$ and $n$ are rather moderate for frequencies in the visible region. Using the data shown in figure~\ref{fig:optix} one obtains an almost constant low reflectance $R\approx 0.2$ over the whole visible region. Since the extinction coefficient is also almost constant in the visible region and it has a quite considerable value, the material $\mgtih$ has a dark, colourless, i.e. black appearance.

Using the extinction coefficient of the $\mgtih$ SQSs shown in figure~\ref{fig:optix} to calculate the absorption $\alpha(\omega) = 2\kappa(\omega) \omega /c$, we obtain a good qualitative agreement with experimental data \cite{borsa2007soa}. The quantitative difference between the calculated and experimental absorption is $\sim 30$~\%, which reflects both the approximations made in
the calculations, and in the extraction of the experimental values.


\section{Summary and conclusions}

$\mgh$ has a high hydrogen storage capacity, but it suffers from poor (de)hydrogenation kinetics and a high thermodynamic stability, which make it unsuitable as a hydrogen storage material. Adding Ti to Mg gives an alloy hydride $\mgtih$ with much faster (de)hydrogenation kinetics for $x\lesssim 0.8$. In this paper we have studied the structure and stability of $\mgtih$, $x= 0$-$1$, by first-principles calculations at the level of density functional theory.

As the most stable structures (at ambient conditions) of $\mgh$ and $\tih$ are rutile and fluorite, respectively, we model such structures for the alloy hydride $\mgtih$. A series of simple, ordered fluorite structures is constructed analogous to the high pressure Mg$_7$TiH$_{16}$ phase, and rutile structures are modelled in $\mgh$ supercells. All cell parameters and atomic positions are optimized and care is taken to converge the total energies with respect to computational parameters such as the basis set size and the $k$-point sampling grid. Upon optimization, all structures stay quite close to either fluorite or rutile. To model disorder in the positions of the metal atoms, we construct a series of SQSs of $\mgtih$ starting from fluorite and rutile structures. Upon optimization the fluorite structures are stable, but in compositions with a high Mg content, such as $\compE$, the structure around the Mg atoms starts to resemble that of the high pressure phase $\beta$-$\mgh$. The SQS rutile structures of $\mgtih$ are unstable for $x\leq 0.875$. From the formation energies of the ordered, as well as the disordered structures, we give evidence for a fluorite to rutile phase transition at a critical composition $x_{\rm c}= 0.8$-$0.9$. This correlates with the experimentally observed sharp decrease in (de)hydrogenation rates around this composition.

Thin films of $\mgti$ show a remarkable optical transition from reflecting to absorbing upon hydrogenation, which is reversible upon dehydrogenation. We calculate the optical response of $\mgtih$, $x=0.5$-$0.875$, in the ordered and disordered structures and show that the black absorbing state is an intrinsic property of this material. $\mgtih$ has a peak at the Fermi level for all compositions $x$, which is composed of Ti d states. However, the resulting metallic plasma only plays a minor role in the optical response in the visible range for compositions with a high Mg content, as interband transitions interfere to decrease the extinction coefficient $\kappa$. Moreover, this effect is enhanced by disorder in the positions of the Ti atoms, which easily destroys the metallic plasma and suppresses the optical reflection even more. The contributions of the H atoms to the bands are spread out over a large energy range as a result of the diverse coordination by Ti and Mg atoms in $\mgtih$. Interband transitions then result in an almost constant optical absorption over a large energy range, causing the black appearance of these compounds.
\section*{Acknowledgments}
The authors thank G.~Kresse and J.~Harl (Universit{\"a}t Wien) for the use of the optical packages, and A.~V.~Ruban (Technical University of Denmark, Lyngby) for his help and for making available his program to generate quasi random structures. This work is part of the Sustainable Hydrogen research program of Advanced Chemical Technologies for Sustainability (ACTS) and the Stichting voor Fundamenteel Onderzoek der Materie (FOM). The use of supercomputer facilities is sponsored by the Stichting Nationale Computerfaciliteiten (NCF). These institutions are financially supported by the Nederlandse Organisatie voor Wetenschappelijk Onderzoek (NWO).


\end{document}